 \documentclass[final,3p,times,twocolumn,sort&compress]{elsarticle}

 \usepackage{graphicx,amssymb,amsthm,lineno}
 \usepackage{dcolumn,multirow}

\biboptions{square}

\begin{document}

\begin{frontmatter}



\title{Three-state Potts model on Non-local Directed Small-World Lattices}


\author[ufsa]{Carlos Handrey Araujo Ferraz\corref{cor1}}
\ead{handrey@ufersa.edu.br}
\author[ufsa]{Jos\'e Luiz Sousa Lima}
\ead{jlima@ufersa.edu.br}
\cortext[cor1]{Corresponding author}
\address[ufsa]{Exact and Natural Sciences Center, Universidade Federal Rural do Semi-\'Arido-UFERSA, PO Box 0137, CEP 59625-900, Mossor\'o, RN, Brazil}

\begin{abstract}
In this paper, we study the non-local directed Small-World (NLDSW) disorder effects in the three-state Potts model as a form to capture the essential features shared by real complex systems where non-locality effects play a important role in the behavior of these systems.  Using Monte Carlo techniques and finite-size scaling analysis, we estimate the infinite lattice critical temperatures and the leading critical exponents in this model. In particular, we investigate the first- to second-order phase transition crossover when NLDSW links are inserted.  A cluster-flip algorithm was used to reduce the critical slowing down effect in our simulations. We find that for a NLDSW disorder densities $p<p^{*}=0.05(4)$, the model exhibits a continuous phase transition falling into a new universality class, which continuously depends on the value of $p$, while for $p^{*}\leqslant p \leqslant 1.0$, the model presents a weak first-order phase transition.

\end{abstract}


\end{frontmatter}


\section{Introduction \label{sec:int}}

In the past, the connection topology had been assumed to be either completely random or completely regular. But many biological, technological and social networks lie somewhere between these two extremes. The small-world (SW) topology~\cite{watts} is suitable for this purpose and constitutes an interesting attempt to translate complex networks as physical, biological and social networks into a simple model. Applications to earth sciences, brain sciences, computing and sociology have been extensively reported. Remarkably, dynamical systems with small-world coupling exhibit enhanced signal-propagation speed~\cite{ferraz2006} and synchronizability~\cite{hong2002} when compared to systems with regular coupling. SW networks are obtained by randomly replacing a fraction {\it p} of links of a regular lattice with new random links. As a result of this random rewiring, SW networks interpolate between a regular lattice \({\it p}=0\) and a completely random graphs \({\it p}=1\).

Two different types of SW networks have been purposed to understand the underlying features found in real complex networks: Undirected (standard) SW networks~\cite{watts} and directed SW networks~\cite{sanchez}.  Undirected SW networks are formed by symmetric links in the sense that if a given node {\it X} of the network is linked to a given node {\it Y}, then the node {\it Y} must also be linked to the node {\it X}, such as in the network of movie actors' collaboration or authorship of scientific papers. While directed SW networks are formed by asymmetric links, i.e, if a given node {\it X} of the network is linked to a given node {\it Y}, the node {\it Y} may not be linked to {\it X} but to another. Such examples of this kind of network are World Wide Web, lending transactions and asymmetric synaptic connections.

Although spin-like models on SW topology has been intensity studied in last two decades, there have rarely been studies concerning directed SW topology where only non-local connections take place in these models. Knowing how local and non-local SW directed disorder separately influence the dynamics of systems is particularly important. Moreover, the criticality in systems with this kind of connection disorder remains little known. Therefore, this issue necessitates extensive computational research for accurately estimating the static critical exponents in these systems.

While earlier studies~\cite{fernandes2010, lima2013, silva2013} employing directed SW network in different spin models have showed that this connection disorder can change the universality class of these models. However, they have not distinguishably focused on non-local directed SW disorder. But many systems found in the nature exhibit non-locality effects in which long-range interactions entirely dominate the time evolution of these systems. Such examples are sexually transmitted diseases, genetic recombination and quantum computation, among others.

\begin{figure}[!t]
 \centering
 \includegraphics[scale=0.30]{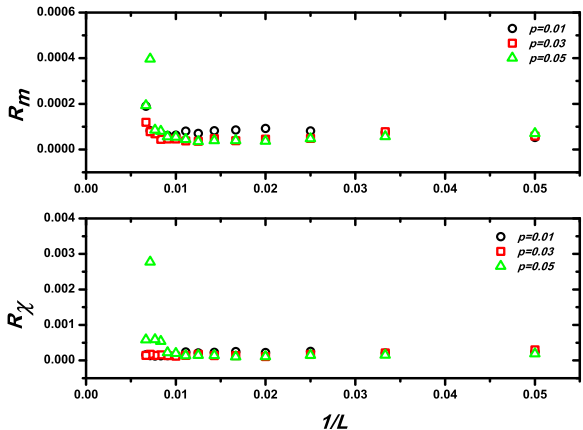}
 \caption{Plot of the relative variance for both the magnetization $R_{m}$ (top panel) and the susceptibility $R_{\chi}$ (bottom panel) versus the inverse of the lattice size at the effective critical temperature $T_{c}(L)$ (as given by Eq.~(\ref{eq:20})) for three studied values of the disorder density $p$.} \label{fig:1a}
\end{figure}

\begin{figure*}[t]
\begin{minipage} [!r]{0.49\linewidth}
\includegraphics*[scale=0.30,angle=0]{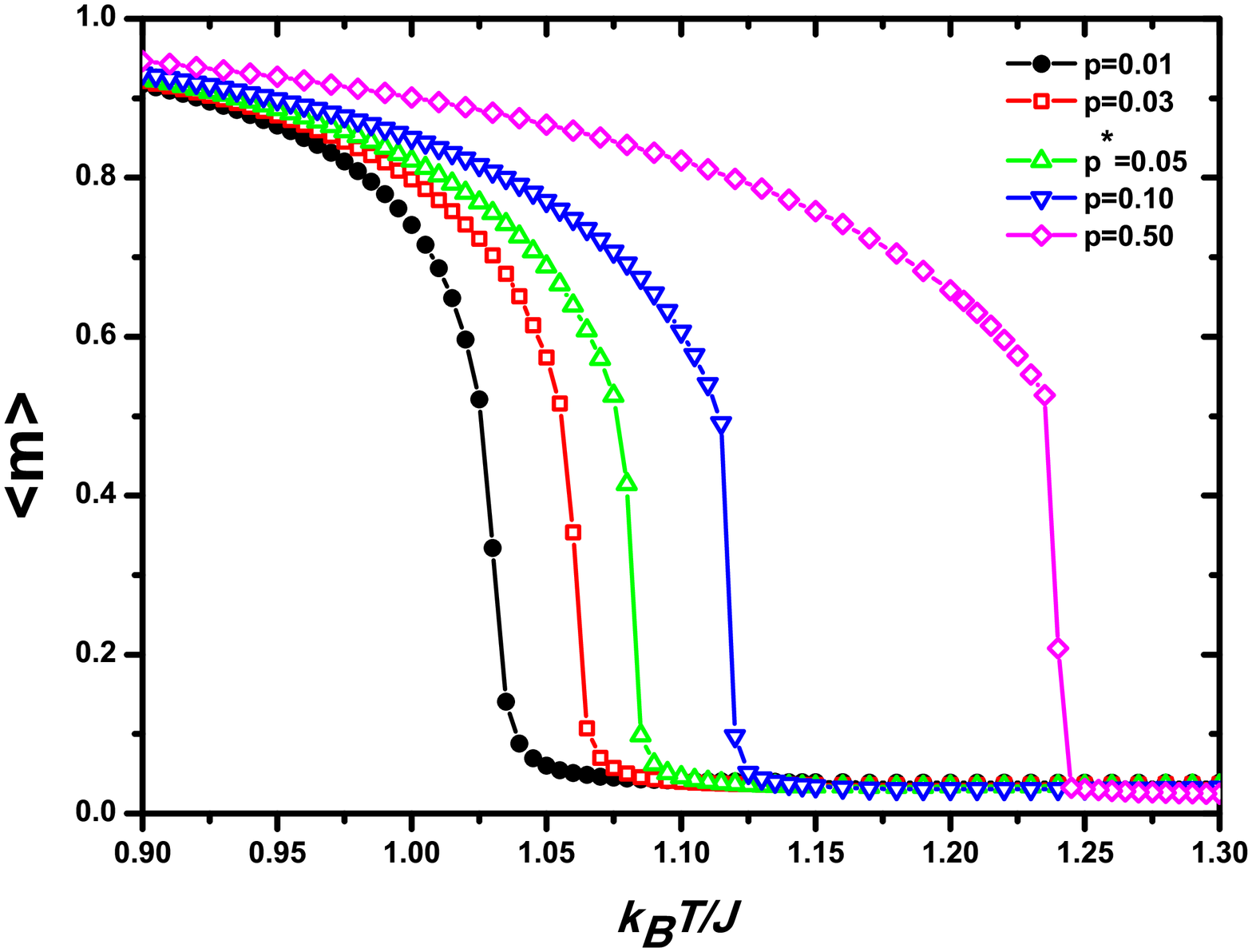}
\caption{Magnetization per spin against temperature $k_{B}T/J$ for $q=3$ Potts model on the $150 \times 150$ NLDSW lattice for $p=0.01$ (black circle), $p=0.03$ (red square), $p^{*}=0.05$ (green triangle), $p=0.10$ (blue upside down triangle) and $p=0.50$ (magenta diamond). A typical second-order phase transition is clear for $p=0.01$, $0.03$ and $0.05$, while for $p=0.10$ and $0.50$, a first-order phase transition takes place.}\label{fig:1}
\end{minipage}\hfill
\begin{minipage}[!r]{0.49\linewidth}
\includegraphics*[scale=0.30,angle=0]{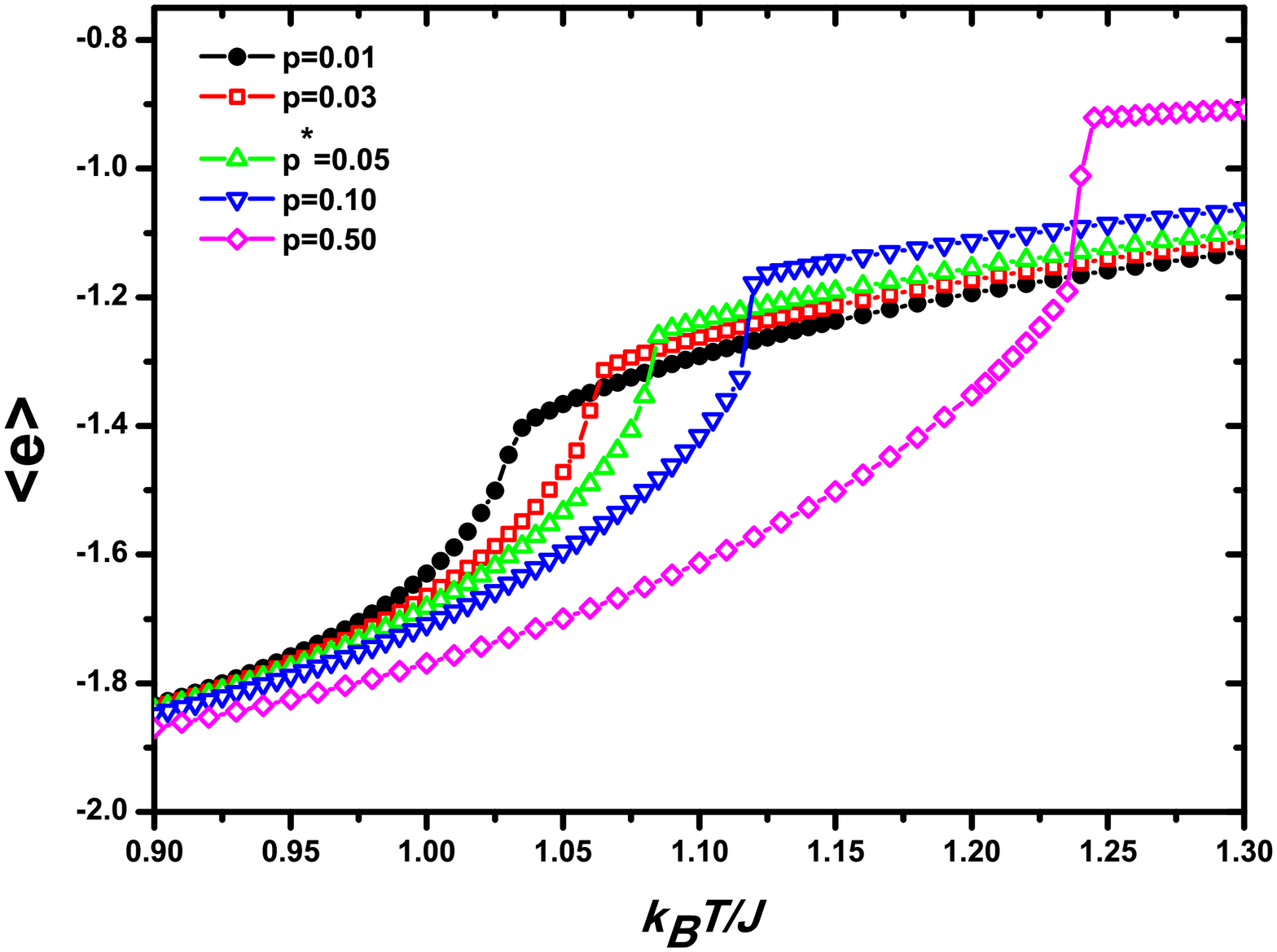}
\caption{Energy per spin against temperature $k_{B}T/J$ for $q=3$ Potts model on the $150 \times 150$ NLDSW lattice for $p=0.01$ (black circle), $p=0.03$ (red square), $p^{*}=0.05$ (green triangle), $p=0.10$ (blue upside down triangle) and $p=0.50$ (magenta diamond). For the cases $p=0.10$  and $p=0.50$, a finite amount of latent heat arises at $T_{c}(L)$ which is given by the discontinuous jump in the energy curve at this point.}\label{fig:2}
\end{minipage}
\end{figure*}

\begin{figure*}[!t]
\begin{minipage} [!r]{0.49\linewidth}
\includegraphics*[scale=0.30,angle=0]{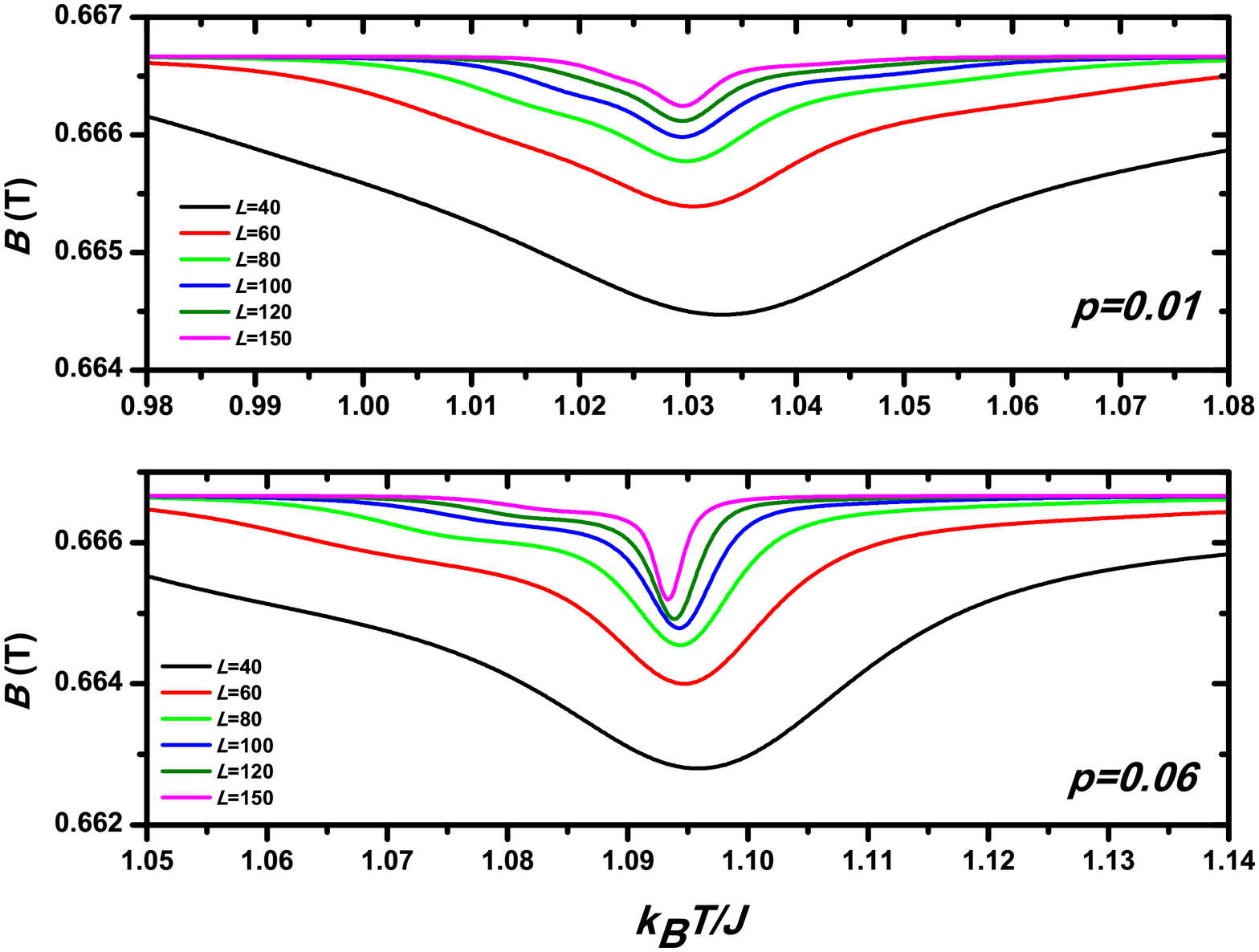}
\caption{The fourth-order energetic cumulant $B$ as a function of the temperature for $p=0.01$ (top panel) and $p=0.06$ (bottom panel) considering several lattice size. In this figure, the curves are obtained by standard histogram reweighting of the simulation data at one given value of temperature.}\label{fig:3a}
\end{minipage}\hfill
\begin{minipage} [!r]{0.49\linewidth}
\includegraphics*[scale=0.30,angle=0]{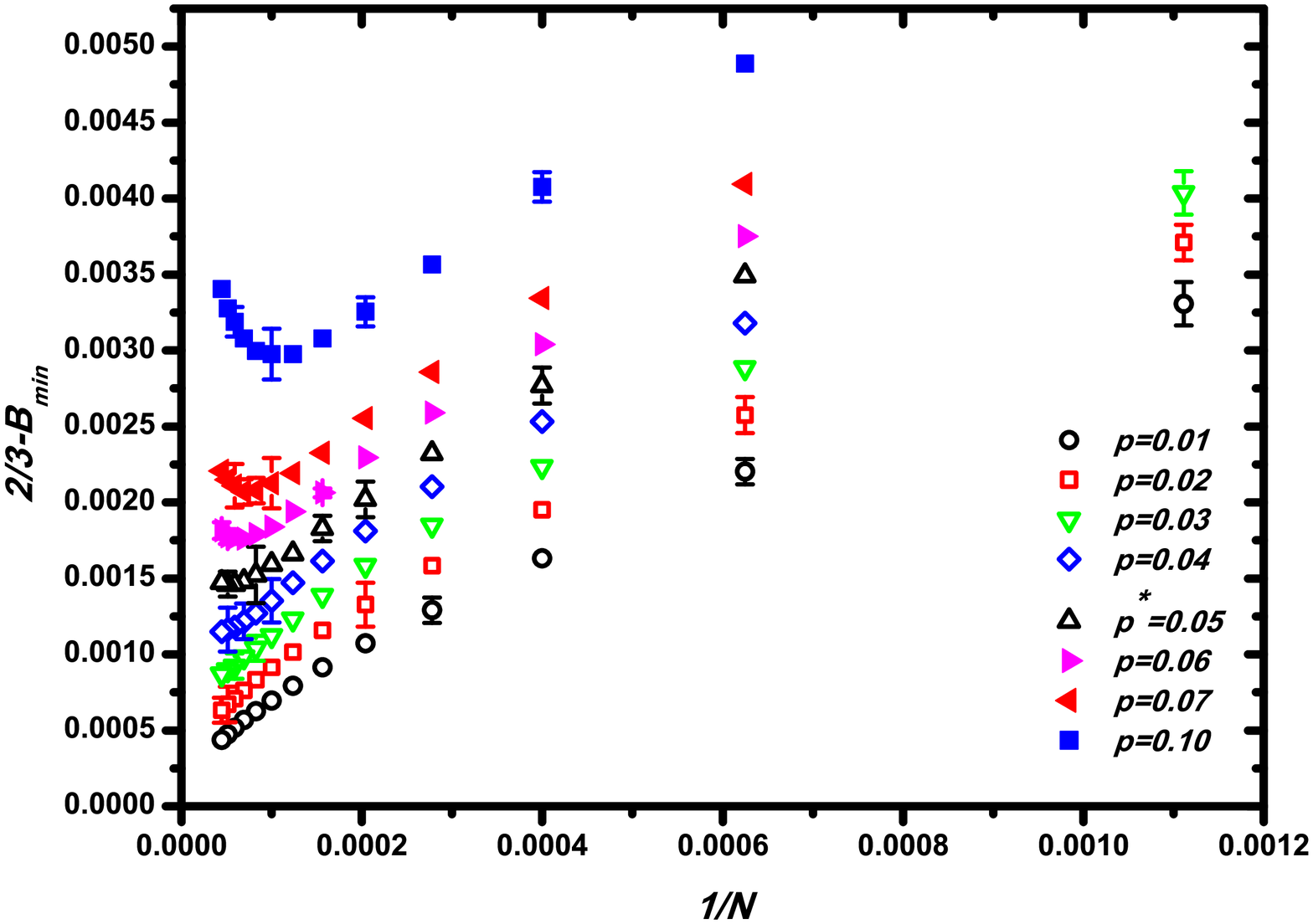}
\caption{Plot of the quantity $(2/3-B_{min})$ as function of $1/N$ for several NLDSW disorder densities $p$.  By extrapolating the data of $(2/3-B_{min})$, we observe a second-order transition for $p\leq 0.05$, since $(2/3-B_{min})\rightarrow 0$ as $1/N\rightarrow 0$, while for $p>0.05$, $(2/3-B_{min})\rightarrow \epsilon \neq 0$, and all transitions are first-order.}\label{fig:3}
\end{minipage}
\end{figure*}

\begin{figure*}[!t]
\begin{minipage}[!r]{0.49\linewidth}
\includegraphics*[scale=0.30,angle=0]{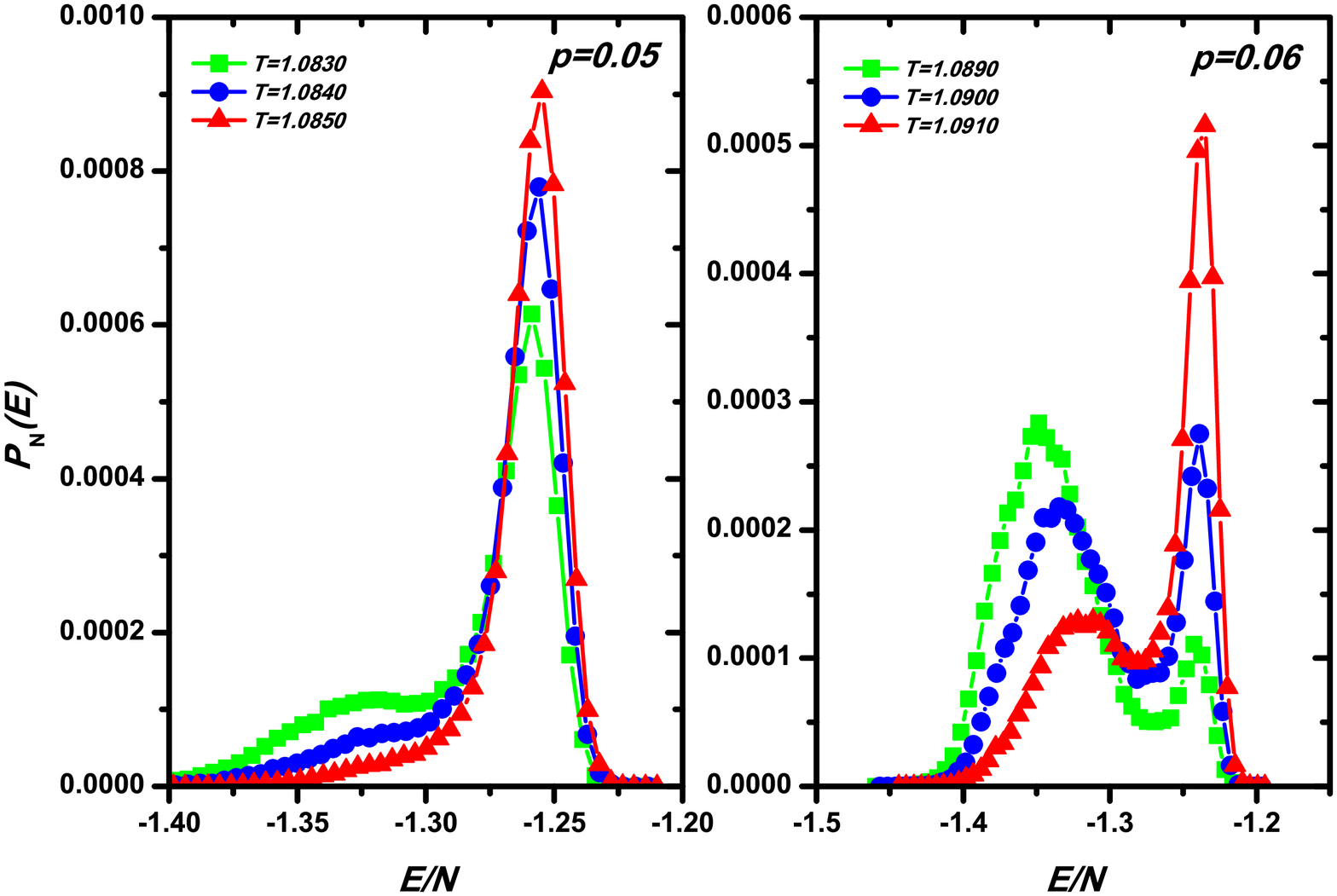}
\caption{Plot of the probability density function (PDF) of the energy $P_{N}(E)$ on $150 \times 150$ NLDSW lattice for $p^{*}=0.05$ (left panel) and $p=0.06$ (right panel) at three different temperatures close to the critical point $T_c$ for each case.}\label{fig:4}
\end{minipage}\hfill
\begin{minipage} [!r]{0.49\linewidth}
\includegraphics*[scale=0.30,angle=0]{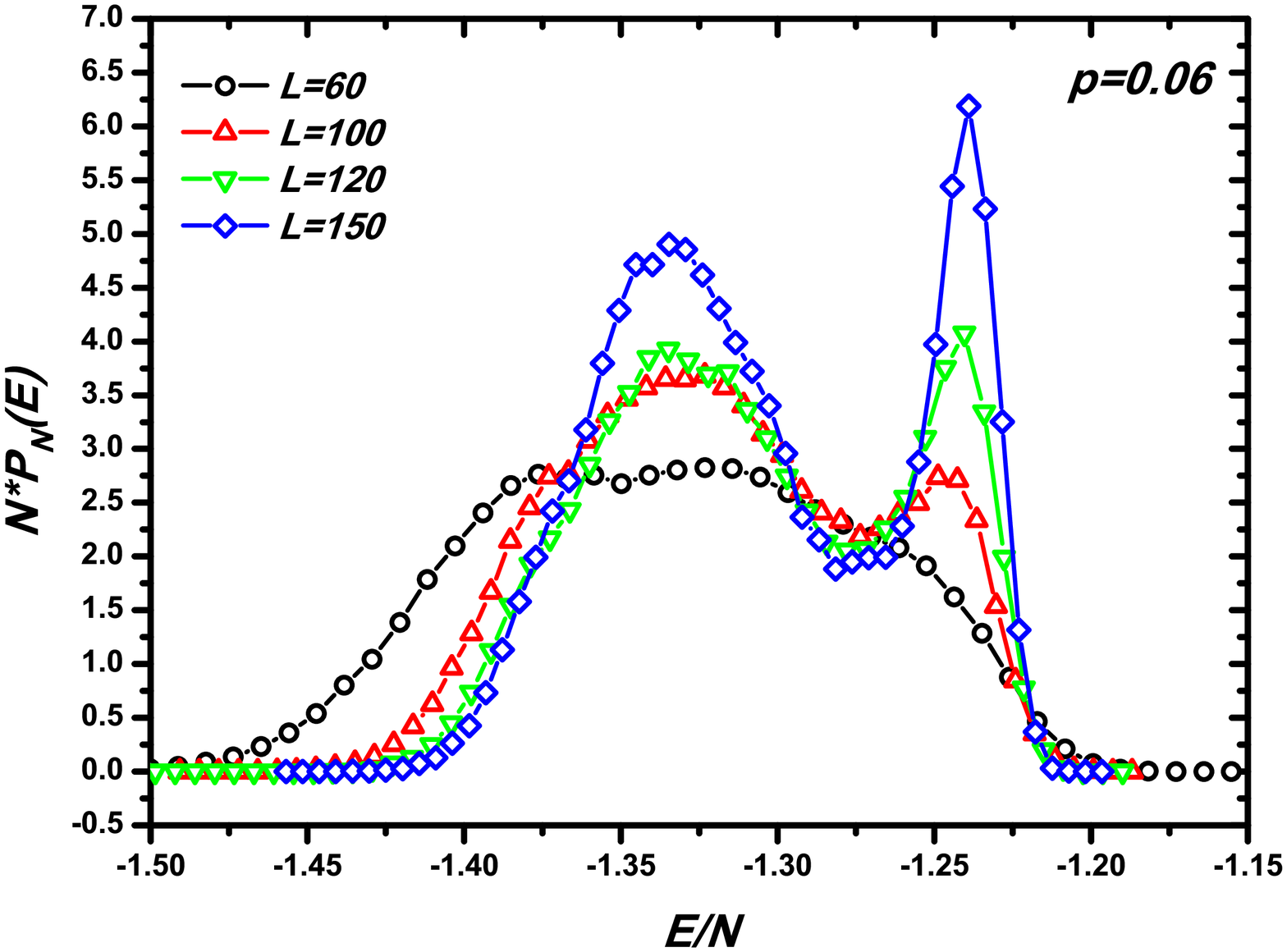}
\caption{Plot of the rescaled probability density function $N P_{N}(E)$ for $p=0.06$ obtained at each effective critical temperatures $T_{c}(L)$ considering several lattice size. The double-peak structure in the PDF is evident in all cases.}\label{fig:4a}
\end{minipage}
\end{figure*}

\begin{figure*}[t]
\begin{minipage} [!r]{0.49\linewidth}
\includegraphics*[scale=0.30,angle=0]{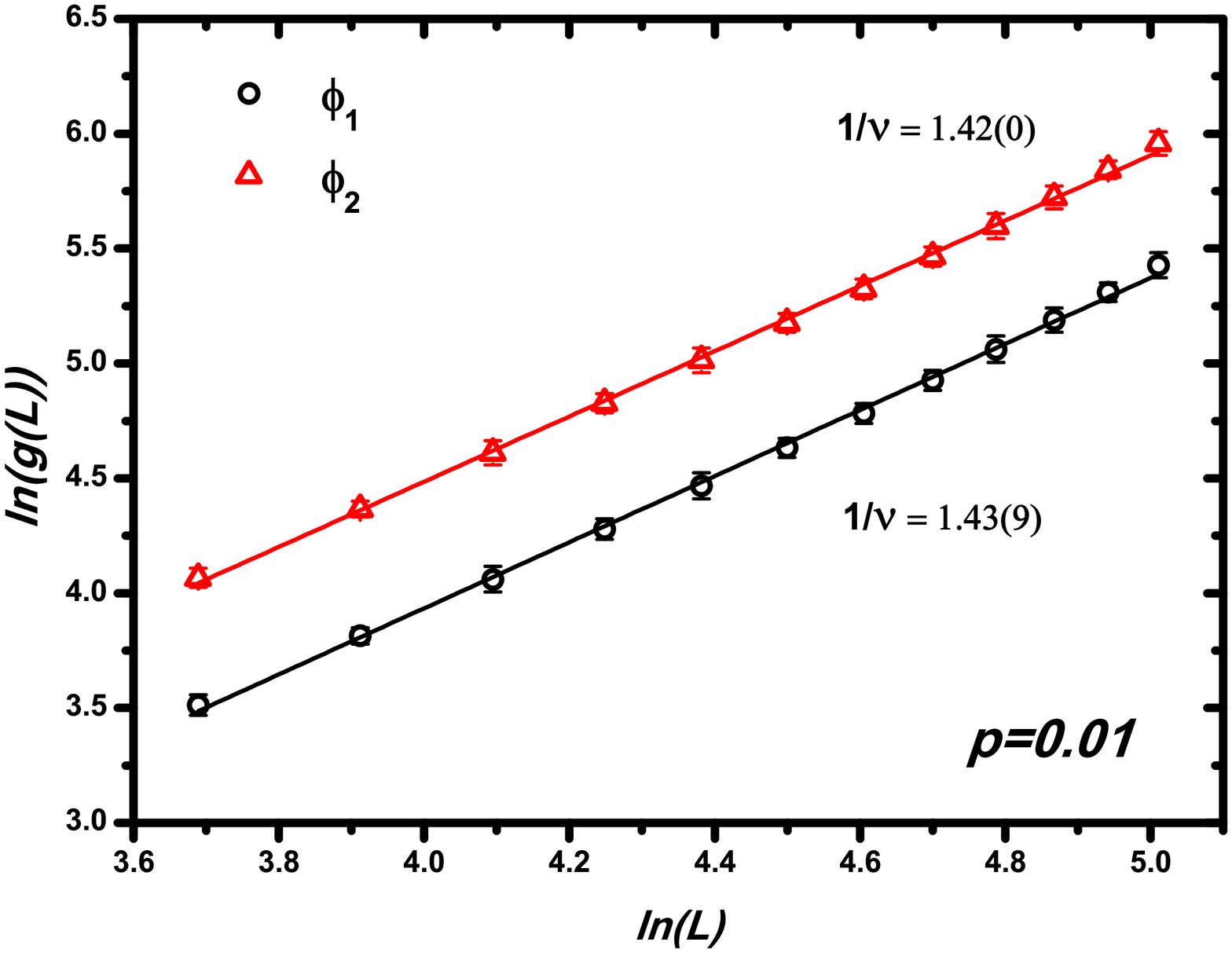}
\caption{Log-log plot of the size dependence of the maximum values of the thermodynamic derivatives $g(L) \equiv \phi_1$ (black circle) and $\phi_2$ (red triangle) for the $q=3$ Potts model on the NLDSW lattice for the case $p=0.01$.}\label{fig:5}
\end{minipage}\hfill
\begin{minipage}[!r]{0.49\linewidth}
\includegraphics*[scale=0.30,angle=0]{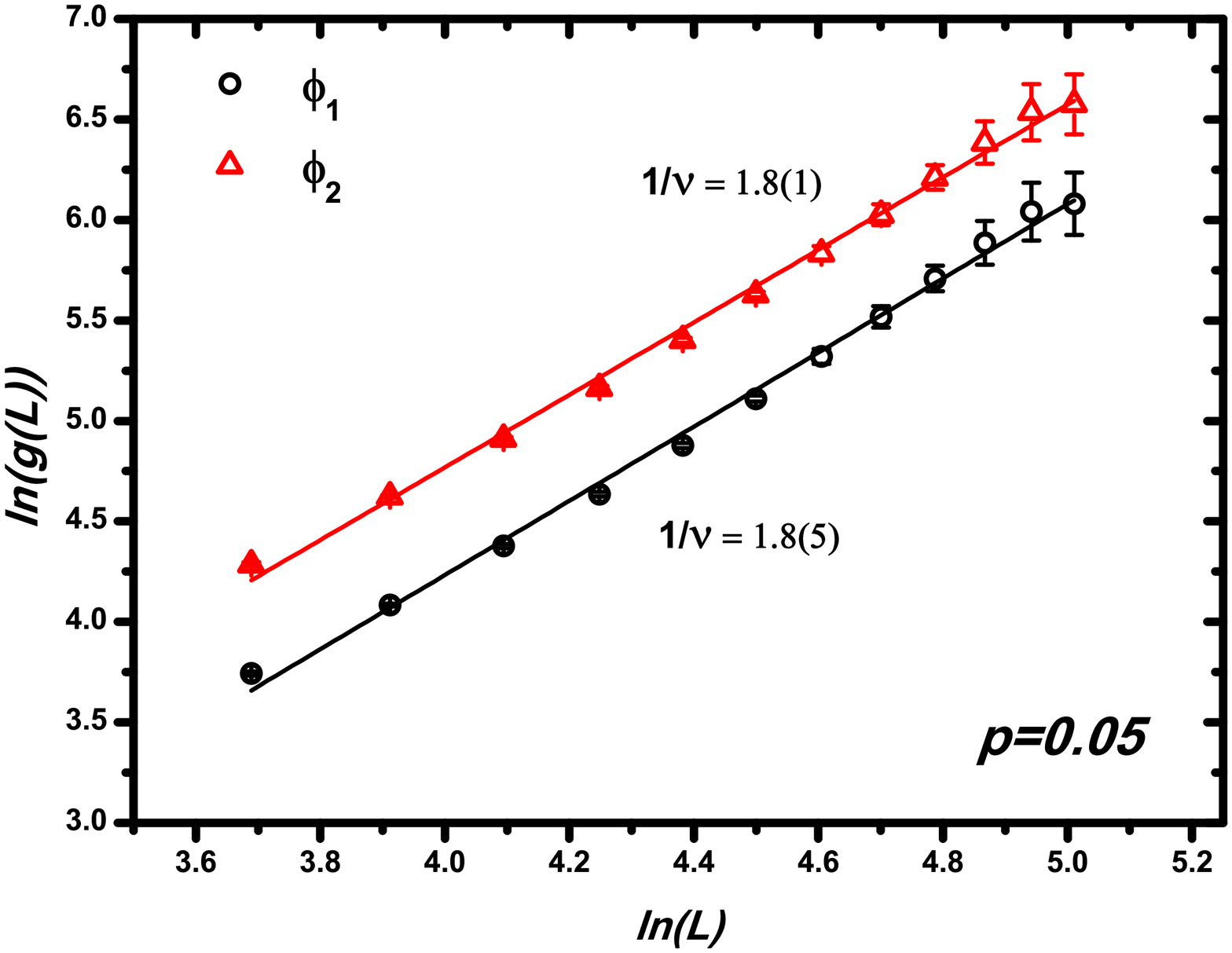}
\caption{Log-log plot of the size dependence of the maximum values of the thermodynamic derivatives $g(L) \equiv \phi_1$ (black circle) and $\phi_2$ (red triangle) for the $q=3$ Potts model on the NLDSW lattice for the case $p=0.05$. }\label{fig:6}
\end{minipage}
\end{figure*}

\begin{figure*}[!t]
\begin{minipage} [!r]{0.49\linewidth}
\includegraphics*[scale=0.30,angle=0]{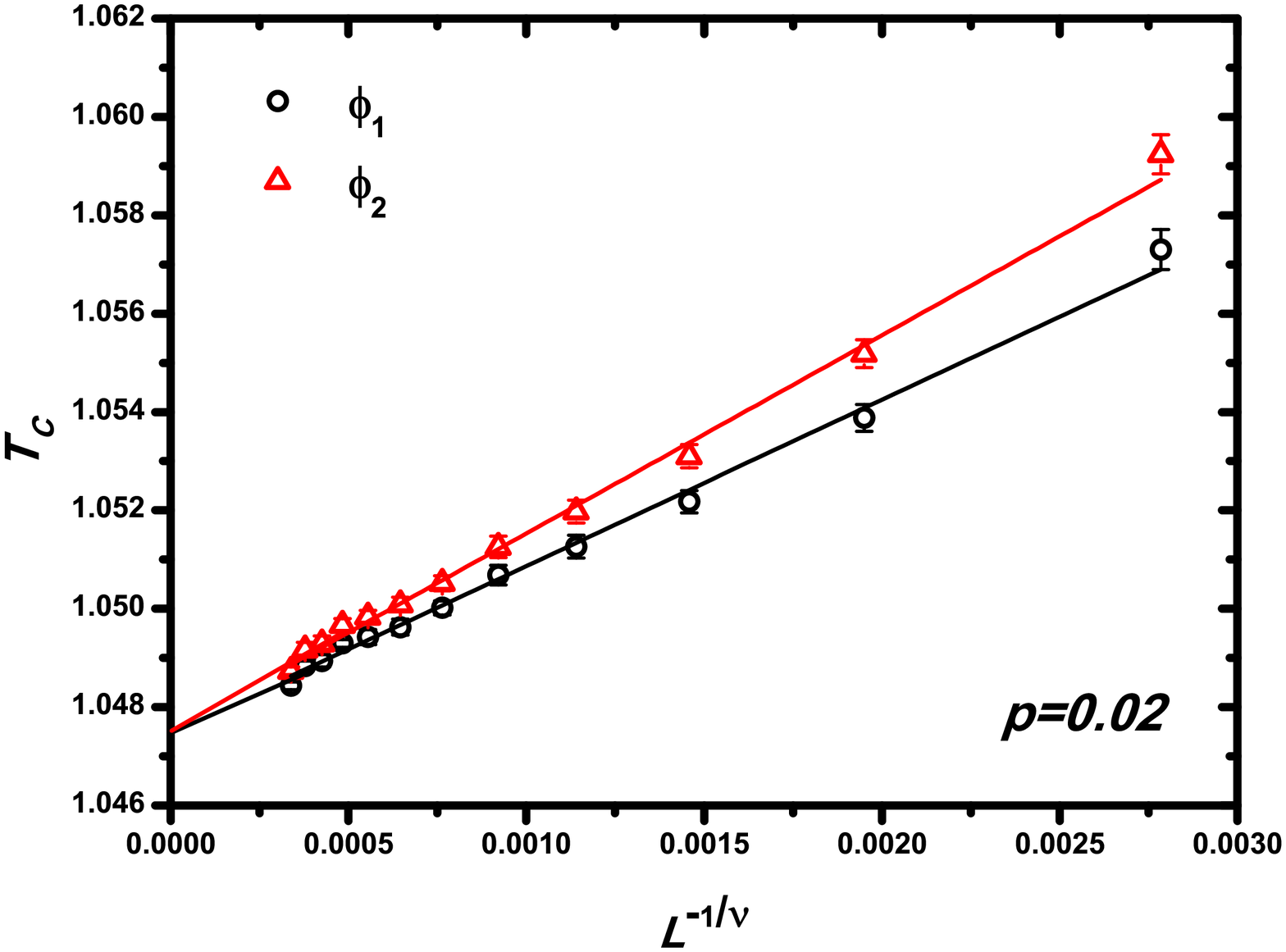}
\caption{Size dependence of the effective critical temperatures $T_c(L)$ for the case $p=0.02$ obtained from the location of the maximum values of the thermodynamics derivatives $\phi_1$ and $\phi_2$ . The curves are straight line fits to Eq.~\ref{eq:20} with $\nu=0.62(6)$. }\label{fig:7}
\end{minipage}\hfill
\begin{minipage}[!r]{0.49\linewidth}
\includegraphics*[scale=0.30,angle=0]{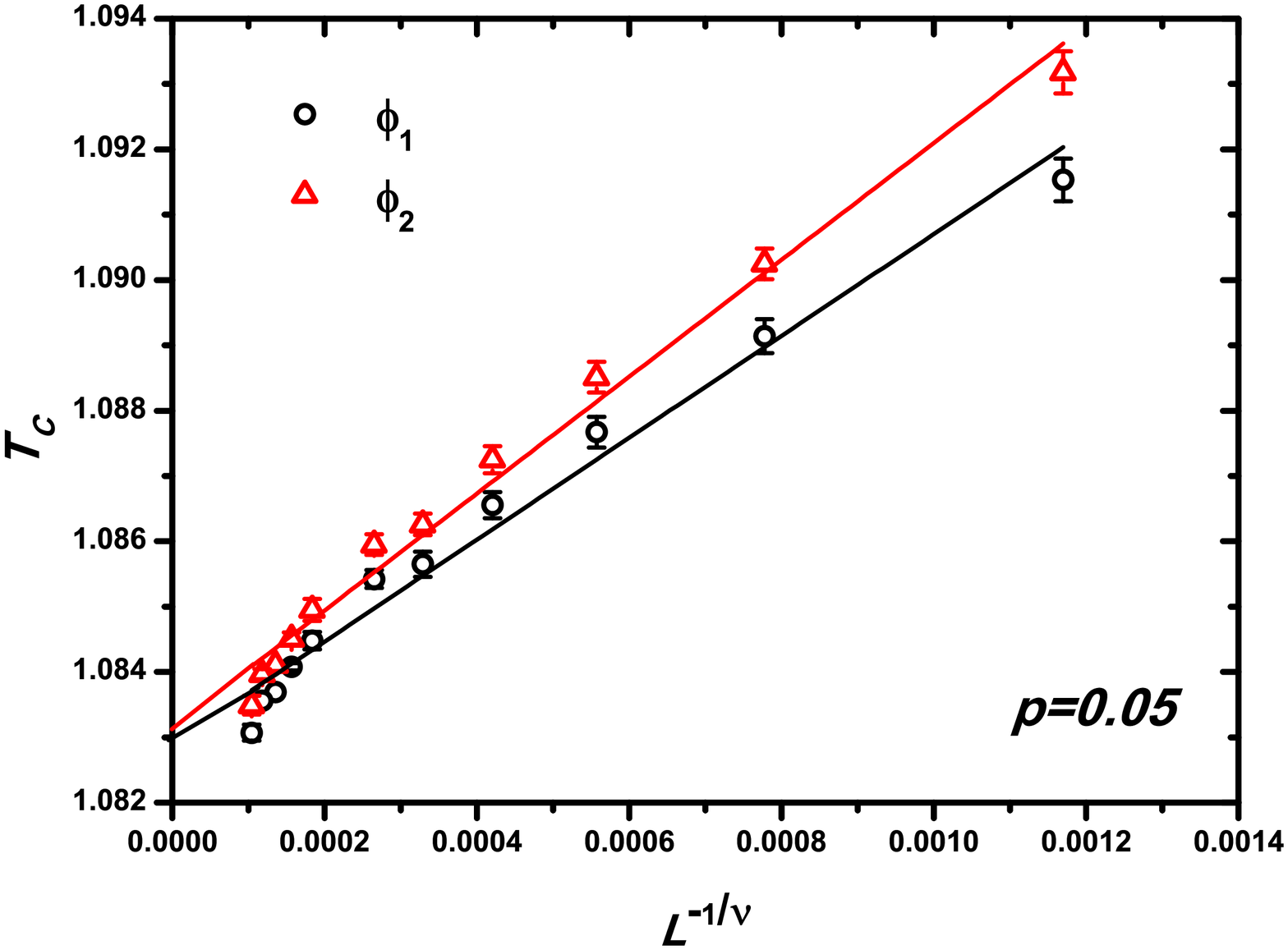}
\caption{Size dependence of the effective critical temperatures $T_c(L)$ for the case $p=0.05$ obtained from the location of the maximum values of the thermodynamics derivatives $\phi_1$ and $\phi_2$ . The curves are straight line fits to Eq.~\ref{eq:20} with $\nu=0.54(8)$.}\label{fig:8}
\end{minipage}
\end{figure*}

\begin{figure*}[t]
\begin{minipage} [!r]{0.49\linewidth}
\includegraphics*[scale=0.30,angle=0]{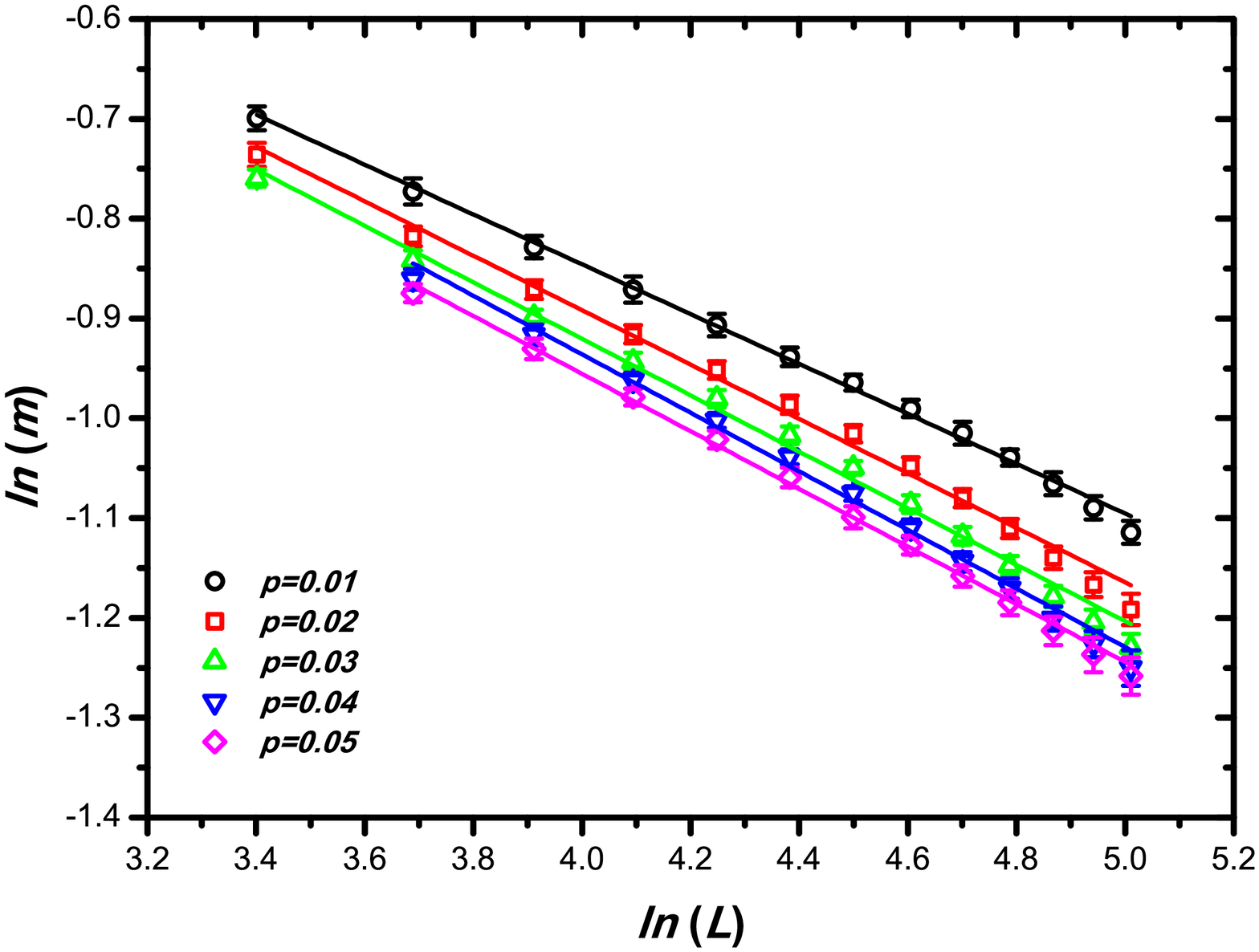}
\caption{Log-log plot of the magnetization $m$ (measured at the temperature with maximum value of $dm/dK$) versus linear size $L=\sqrt{N}$ for $q=3$ Potts model on NLDSW lattices for $p=0.01$ (black circle), $p=0.02$ (red square), $p=0.03$ (green triangle), $p=0.04$ (blue upside down triangle) and $p=0.05$ (magenta diamond). }\label{fig:9}
\end{minipage}\hfill
\begin{minipage}[!r]{0.49\linewidth}
\includegraphics*[scale=0.30,angle=0]{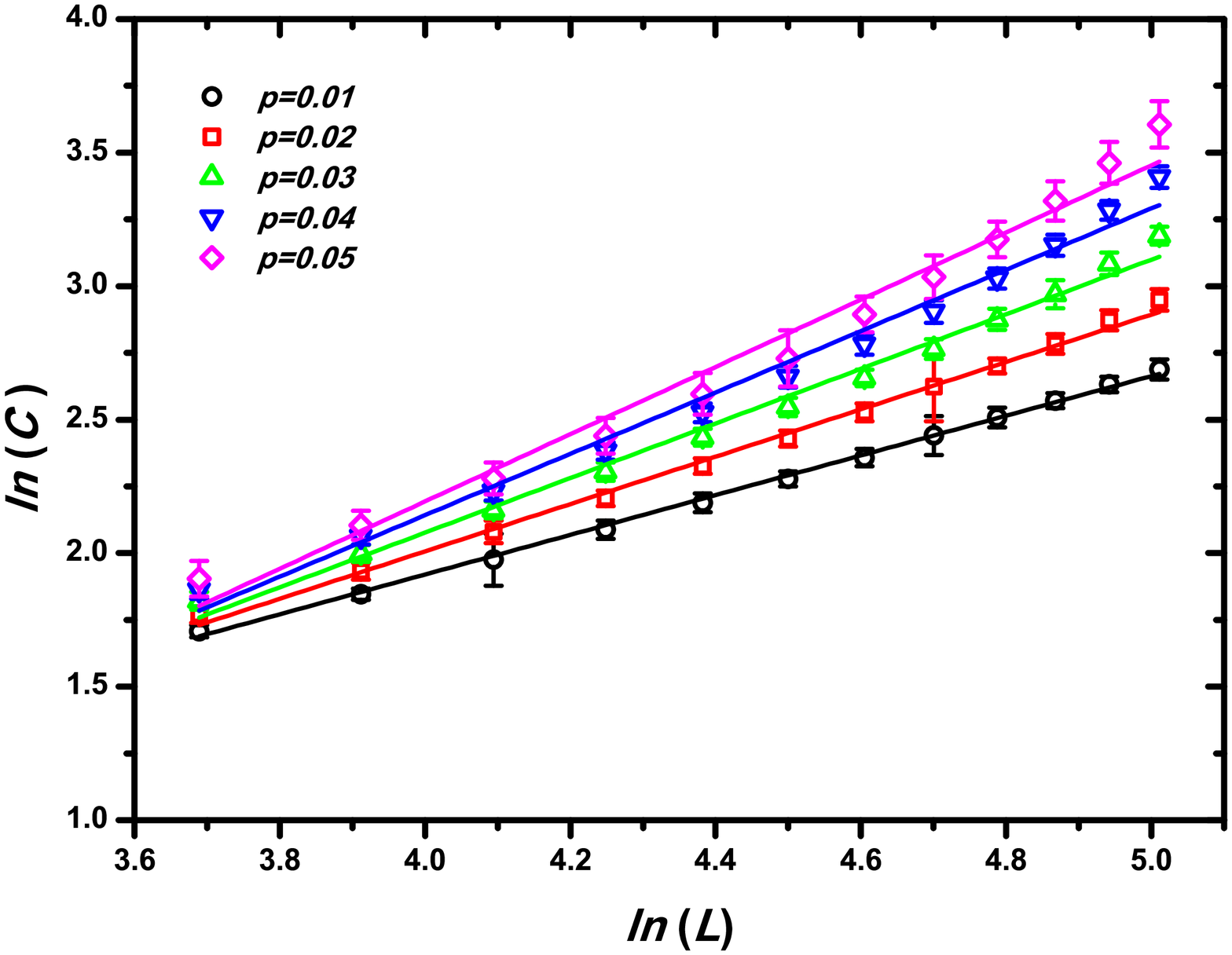}
\caption{ Log-log plot of the maximum values of $C$ versus linear size $L=\sqrt{N}$ for $q=3$ Potts model on NLDSW lattices for $p=0.01$ (black circle), $p=0.02$ (red square), $p=0.03$ (green triangle), $p=0.04$ (blue upside down triangle) and $p=0.05$ (magenta diamond). }\label{fig:10}
\end{minipage}
\end{figure*}

\begin{figure*}[!t]
\begin{minipage} [!r]{0.49\linewidth}
\includegraphics*[scale=0.30,angle=0]{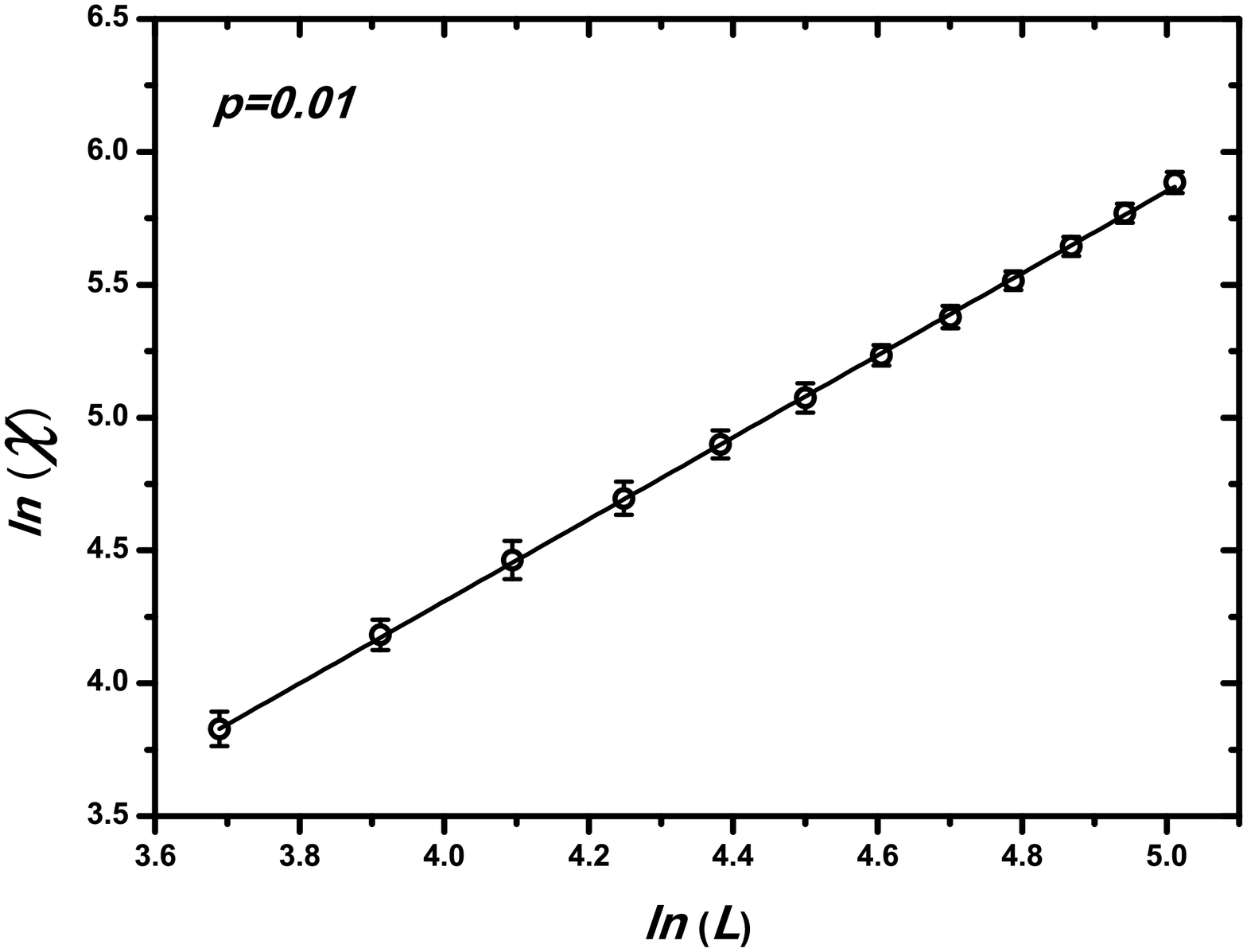}
\caption{Log-log plot of the maximum values of $\chi$ versus linear size $L=\sqrt{N}$ for the $q=3$ Potts model on the NLDSW lattice for the case $p=0.01$.}\label{fig:11}
\end{minipage}\hfill
\begin{minipage}[!r]{0.49\linewidth}
\includegraphics*[scale=0.30,angle=0]{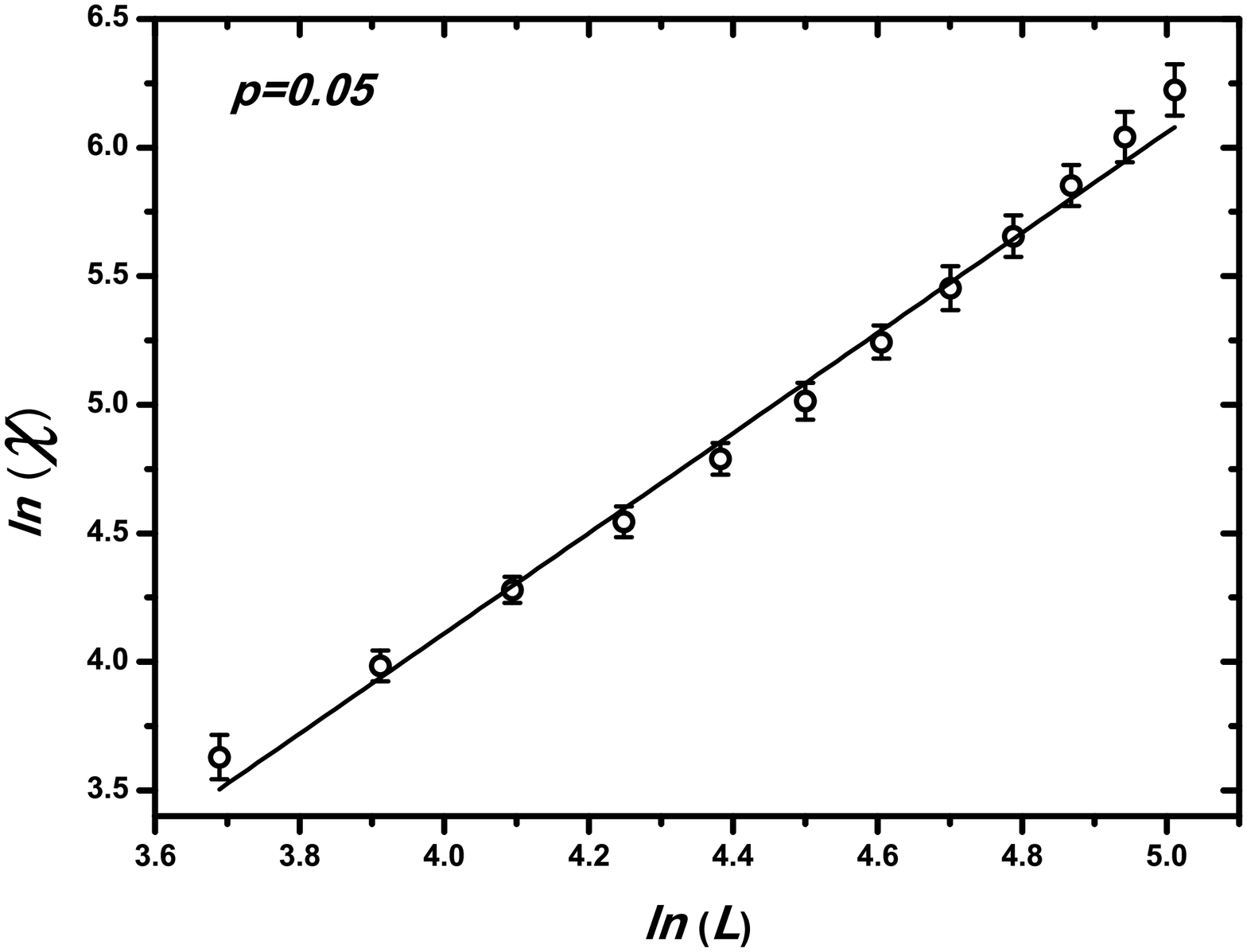}
\caption{Log-log plot of the maximum values of $\chi$ versus linear size $L=\sqrt{N}$ for the $q=3$ Potts model on the NLDSW lattice for the case $p=0.05$.}\label{fig:12}
\end{minipage}
\end{figure*}

In this paper we study the effects of the non-local directed Small-World (NLDSW) disorder in the three-state Potts model as form to capture the essential features shared by real complex systems where non-locality effects play a important role in the behavior of these systems. We choose the ferromagnetic Potts model~\cite{wu82,binder81a} because its simplicity and well-known phase transition properties. Using Monte Carlo (MC) techniques~\cite{murta2014,murta2012,jin2006} and finite-size scaling (FSS) analysis, we estimate the infinite lattice critical temperatures and the leading critical exponents in this model. In particular, we investigate the first- to second-order phase transition crossover when non-local directed links are inserted.  In order to reduce the critical slowing down effect~\cite{heermann88}, we used a cluster-flip algorithm to evolve the lattices studied over time. Periodic boundary conditions were also used to avoid the boundary effects caused by the finite size. In this study, we make an analysis of several thermodynamic quantities including the specific heat and susceptibility as well as the fourth-order energetic cumulant, derivatives and logarithmic derivatives of the magnetization.

The contents of the article are organized as follow. In section \ref{sec:mms}, we describe details of the model and Monte Carlo simulation background. In section \ref{sec:r}, we  present and discuss the results. Finally, in section \ref{sec:c}, we make the conclusions.

\section{ Model and Monte Carlo Simulation \label{sec:mms}}

The NLDSW lattices used in MC simulations were constructed in similar way as in S\'anchez {\it el al}~\cite{sanchez}. First, we start from a regular square lattice consisting of sites linked to their four nearest neighbors by both outgoing and incoming links. Then, with probability $p$, we reconnect every nearest-neighbor outgoing link to a new site randomly chosen provided that it is neither the site itself (self-interaction) nor any of its four nearest neighbor (local interaction). After repeating this procedure for every outgoing link, a new lattice is constructed with a density $p$ of NLDSW links. In this lattice, each site will have four outgoing links but a random number of incoming links.
\begin{table}[!b] \small
\caption{\label{tab:1} Estimates of the critical temperatures $T_c$ and reciprocal correlation-length exponents $1/\nu$ for three-state Potts model on NLDSW lattices with different disorder densities $p$.}
\centering
\begin{tabular}{ccccc}
\hline \hline
 & & & &  \\
$p$ &  $T_c(\phi_1)$ & $T_c(\phi_2)$ & $1/\nu(\phi_1)$ & $1/\nu(\phi_2)$  \\ \\ \hline
 & & & &  \\
$0.01$&$1.029(4)$&$1.029(4)$&$1.439\pm 0.014$&$1.420\pm 0.013$ \\
$0.02$&$1.047(5)$&$1.047(5)$&$1.610\pm 0.030$&$1.583\pm 0.030$ \\
$0.03$&$1.061(3)$&$1.061(5)$&$1.740\pm 0.040$&$1.710\pm 0.040$ \\
$0.04$&$1.072(1)$&$1.072(2)$&$1.850\pm 0.050$&$1.830\pm 0.050$ \\
$0.05$&$1.082(9)$&$1.083(2)$&$1.845\pm 0.037$&$1.810\pm 0.035$ \\
$0.054$&$-$&$-$&$2.062\pm 0.066$&$2.010\pm 0.062$ \\
$0.06$&$-$&$-$&$2.118\pm 0.068$&$2.062\pm 0.063$ \\
$0.07$&$-$&$-$&$2.110\pm 0.073$&$2.051\pm 0.068$ \\
\hline \hline
 & & & &
\end{tabular}
\end{table}
To study the critical behavior in NLDSW lattices, we use a Wolff algorithm~\cite{wolff} to update the lattices studied. For a fixed temperature, we define a Monte Carlo step (MCS) per spin by accumulating the flip times of all the spins and then dividing them up by the total spin number. The Hamiltonian of the {\it q}-states ferromagnetic Potts model ($J>0$) can be written as
\begin{equation}\label{eq:7}
H =  - J\sum\limits_{ < i,j > } {\delta (\sigma _i ,} {\kern 1pt} {\kern 1pt} \sigma _j),
\end{equation}
where $\delta$ is the Kronecker delta function, $J$ is the exchange coupling, and the sum runs over all nearest neighbors of $\sigma_{i}$.  The temperature can be defined as $T=J/k_{B}K$, where $k_{B}$ is the Boltzmann constant. We also define the order parameter $m$ as
\begin{equation}\label{eq:9}
m = \frac{1}
{{(q - 1)}}(N_{max}q L^{-d}  - 1)
\end{equation}
and the energy per spin as $e=E/N$, where $N_{max}$ is the maximum number of spins in the same state, $L^{d}=N$ is the total number of spins, and $d$ is the lattice dimension. In this study, $q=3$ and $d=2$. Once the critical region is established, we apply the single histogram method~\cite{ferrenberg91,ferrenberg95} along with FSS analysis to obtain reliable estimates of the critical temperature and critical exponents.
System sizes up to $N=22500$ are used in these simulations with $1.5\times10^6$  MCS per spin performed at a given temperature $T_{0}$, where $5\times10^5$ configurations are discarded for thermalization. For each system size considered, we averaged over $100$ ($N=1600$ to $N=10000$) and over $50$ independent realizations ($N=12100$ to $N=22500$) to estimate the errors due both to the intrinsic statistical fluctuations and the connectivity disorder.

In order to check the data dispersion due to the disorder averaging procedure when the system achieves its asymptotic regime, we calculated the size dependence of the relative variance for both the magnetization and susceptibility, i.e, $R_{m}=\overline{m^{2}(L)}-{\overline{m(L)}}^2/{\overline{m(L)}}^2$ and $R_{\chi}=\overline{\chi^{2}(L)}-{\overline{\chi(L)}}^2/{\overline{\chi(L)}}^2$, respectively. Fig.~\ref{fig:1a} shows these quantities calculated at their according effective critical temperatures $T_{c}(L)$ (see Eq.~(\ref{eq:20}) below) versus the inverse of lattice size for three studied values of p. From these figure, it is seen that the error on the estimates of both the magnetization and the susceptibility approaches a finite value when $L$ increases.

The static thermodynamics quantities such as specific heat, magnetic susceptibility, logarithmic derivatives of the order parameter, and Binder's fourth-order cumulants~\cite{binder81,challa86} are calculated inside the critical region and depending on the analysis of the location of the maximum values of these quantities and their magnitudes, one can estimate both the infinite lattice critical temperature and critical exponents. From the fluctuations of the $e$ measurements, we can calculate the specific heat

\begin{equation}\label{eq:10}
C(T) = \frac{K^2}{N}( < e^2  >  -  < e > ^2 ),
\end{equation}
and the fourth-order energetic cumulant
\begin{equation}\label{eq:13a}
B(T) = 1 - \frac{{ < e^4  > }}
{{3 < e^2  > ^2 }}.
\end{equation}
 Similarly, from the fluctuations of $m$, we can calculate the magnetic susceptibility
\begin{equation}\label{eq:12}
\chi (T) = KN( < m^2  >  -  < m > ^2 ),
\end{equation}

We can also calculate the logarithmic derivative of $n$-power of $m$, i.e,
\begin{equation}\label{eq:15}
  \label{eq:15}
 \phi_n = \frac{\partial}{\partial K}\,ln<m^{n}> = \frac{<m^{n}|\,e>}{<m^{n}>}-<e>.
\end{equation}

According to the FSS theory~\cite{fisher,fisher72}, the free energy of a system of linear dimensional $L$ is described by the scaling ansatz
\begin{equation}\label{eq:16}
f(t,h) = L^{ -d } \widetilde {f}(tL^{1/\nu } ,hL^{(\gamma  + \beta )/\nu } ),
\end{equation}
where $t=(T-T_c)/T_c$ ($T_c$ is the infinite lattice critical temperature) and $h$ is the magnetic field.
The leading critical exponents $\alpha$, $\beta$, $\gamma$ and $\nu$ define the universality class of the system.
Considering zero-field regime, the derivatives of Eq.~(\ref{eq:16}) yield important scaling equations, i.e.,
\begin{eqnarray}
  m &=& L^{ - \beta /\nu } \widetilde m(x), \label{eq:17} \\
  C &=& L^{\alpha /\nu } \widetilde C(x), \label{eq:18}\\
  \chi  &=& L^{\gamma /\nu } \widetilde \chi (x), \label{eq:19}
\end{eqnarray}
where $\widetilde m$, $\widetilde C$ and $\widetilde \chi$ are scaling functions, and $x=tL^{1/\nu}$ is the temperature scaling variable. In addition, the critical temperature scales as
\begin{equation}\label{eq:20}
    T_{c}(L)=T_c+a L^{-1/\nu},
\end{equation}
where $a$ is a constant and $T_{c}(L)$ is the effective transition temperature for the lattice of linear size $L$.
This effective temperature can be obtained by the location of the peaks of the above quantities: $\phi_n$, $dU/dK$, $C$ and $\chi$. For first-order two-dimensional transitions, the power-law scaling behavior of the above quantities is expected to diverge as $L^{2}$~\cite{binder84}.

\section{\label{sec:r} Results and Discussion}

\begin{table*}[t]
\caption{\label{tab:2} Estimates of the ratios of the leading critical exponents for three-state Potts model on NLDSW lattices with different disorder densities $p$.}
\centering
\small\addtolength{\tabcolsep}{15pt}
\begin{tabular}{ccccc}
\hline \hline
 & & & &  \\
 & $\nu$ & $\alpha/\nu$ & $\beta/\nu$ & $\gamma/\nu$  \\ \\ \hline
 & & & &  \\
$p=0.01$&$0.70\pm 0.01$&$0.714\pm 0.013$&$0.249\pm 0.004$&$1.554\pm 0.007$ \\
$p=0.02$&$0.63\pm 0.01$&$0.846\pm 0.020$&$0.273\pm 0.007$&$1.639\pm 0.022$ \\
$p=0.03$&$0.58\pm 0.01$&$0.966\pm 0.031$&$0.283\pm 0.007$&$1.736\pm 0.039$ \\
$p=0.04$&$0.54\pm 0.01$&$1.150\pm 0.040$&$0.293\pm 0.008$&$1.852\pm 0.046$ \\
$p=0.05$&$0.55\pm 0.01$&$1.282\pm 0.057$&$0.289\pm 0.005$&$1.965\pm 0.056$ \\
\hline \hline
 & & & &
\end{tabular}
\end{table*}

In order to determine both the critical region and the order of the phase transition in the  $q=3$-Potts model on NLDSW lattices, we calculated the order parameter $m$ and energy per spin $e$  for several NLDSW disorder densities $p$ in a wide range of temperature. Figs.~\ref{fig:1} and \ref{fig:2} show the order parameter and energy per spin, respectively, for five different $p$ values. Each data point was averaged over 50 different runs. As one can see, a typical second-order phase transition is clear for $p=0.01,0.03$ and $0.05$, while for $p=0.10$ and $0.50$, a first-order phase transition takes place. Furthermore, for the cases $p=0.10$  and $p=0.50$ a finite amount of latent heat arises at $T_{c}$ which can be calculated by the discontinuous jump $\lambda=\epsilon^{-}_{c}-\epsilon^{+}_{c}$ in the energy curve (Fig.~\ref{fig:2}), where $\epsilon^{-}_{c}$ and $\epsilon^{+}_{c}$ are, respectively, the leftmost and rightmost energy values from the effective critical point $T_{c}(L)$. At $T_{c}(L)$, the critical energy is roughly given by $\epsilon_{c}=(\epsilon^{-}_{c}+\epsilon^{+}_{c})/2$.

Fig.~\ref{fig:3a} shows the fourth-order energetic cumulant $B$ given by Eq.~(\ref{eq:13a}) as a function of the temperature for $p=0.01$ and $p=0.06$ considering several lattice size. In this figure, the curves are obtained by standard histogram reweighting of the simulation data at one given value of temperature.
A more detailed view of the first- to second-order phase transition crossover occurring as $p$ decreases is shown in Fig.~\ref{fig:3}. In this figure, $B_{min}$ is the minimum value of $B$ and the quantity $(2/3-B_{min})$ is plotted as a function of $1/N$ for different probability $p$.  The finite-size scaling of this quantity can reveal the order of the phase transition. Indeed, it can be shown~\cite{heermann88} that, in the thermodynamics limit, if the system undergoes a second-order transition, the quantity $(2/3-B_{min})=0$, otherwise, if the system undergoes a first-order transition, the quantity $(2/3-B_{min})=\epsilon$, being $\epsilon$ a small positive value. By extrapolating the data of $(2/3-B_{min})$ versus $1/N$, we observe a second-order transition for $p\leq 0.05$, since $(2/3-B_{min})\rightarrow 0$ as $1/N\rightarrow 0$, while for $p>0.05$, $(2/3-B_{min})\rightarrow \epsilon$, and all transitions are first-order.

In addition, the probability density function (PDF) of the energy $P_{N}(E)$ helps to confirm this change in the phase transition order at $p^{*}\sim 0.05(4)$. Fig.~\ref{fig:4} displays typical PDFs of the energy for both $p=0.05$ and $p=0.06$, respectively, at three different temperatures close to $T_{c}$. As one can see, for $p^{*}=0.05$, a characteristic single peak in the PDF is indicative of a continuous phase transition taking place, while for $p=0.06$, a double peak in the PDF, which exhibits two coexisting phases, is a clear evidence of a weak first-order transition~\cite{fernandez92}. While Fig~\ref{fig:4a} illustrates the double-peak structures in the PDFs of the energy for $p=0.06$ obtained at each effective critical temperature $T_{c}(L)$ considering several lattice size.

Taking the slope of the log-log plot of the maximum values of the quantities $g(L)\equiv \phi_1$ and $\phi_2$ versus $L$, two different estimates were obtained for $1/\nu$.  Figs.~\ref{fig:5} and \ref{fig:6} show the log-log plot of these quantities for $p=0.01$ and $p=0.05$, respectively. We obtained for $p=0.01$, $1/\nu=1.439\pm 0.014$( $\phi_1$) and $1/\nu=1.420\pm 0.013$ ($\phi_2$). By combining these results, we get $1/\nu=1.430\pm 0.010$. For $p=0.05$, we obtained $1/\nu=1.845\pm 0.037$( $\phi_1$) and $1/\nu=1.805\pm 0.035$ ($\phi_2$) yielding an average value of $1/\nu=1.825\pm 0.025$.  A Similar analysis was also performed for $p=0.02$, $p=0.03$, $p=0.04$, $p=0.054$, $p=0.06$ and $p=0.07$, which yielded average values of $1\nu=1.595\pm 0.020$ for $p=0.02$, $1/\nu=1.725\pm 0.028$ for $p=0.03$, $1/\nu=1.850\pm 0.041$ for $p=0.04$, $1/\nu=2.036\pm 0.045$ for $p=0.054$, $1/\nu=2.090\pm 0.046$ for $p=0.06$ and $1/\nu=2.081\pm 0.050$ for $p=0.07$.  All these estimates are consistent each other, within the error bars. Furthermore, the estimates for $p=0.06$ and $p=0.07$ show that the thermodynamics quantities $\phi_1$ and $\phi_2$ scale as $L^2$ inside the critical region, helping to confirm the presence of the first-order phase transition for $p>0.05(4)$. After obtaining estimates for $1/\nu$, the infinite lattice critical temperatures were computed by plotting the size dependence of the location of the peaks of $\phi_1$ and $\phi_2$ as given by Eq.~\ref{eq:20}. Figs.~\ref{fig:7} and \ref{fig:8} show the finite-size scaling of the effective transition temperatures for $p=0.02$ and $p=0.05$.  For $p=0.02$, we found $T_c=1.047(5)$ ($\phi_1$ and $\phi_2$) and for $p=0.05$, $T_c=1.082(9)$ ( $\phi_1$) and $T_c=1.083(2)$ ( $\phi_2$).  A similar extrapolating analysis was also performed for $p=0.01$, $p=0.03$ and $p=0.04$. These results along with the estimated values of $1/\nu$ for each $p$ are summarized in Table \ref{tab:1}.

Using Eqs.~(\ref{eq:17}-\ref{eq:19}) for the size dependence of the maximum values of $m$, $C$ and $\chi$, we can estimate $\beta /\nu$, $\alpha /\nu$ and $\gamma /\nu$, respectively. Fig.~\ref{fig:9} shows the log-log plot of $m$ (measured at the temperature with maximum value of $dm/dK$) versus the linear size of the system $L$ for several $p$ values. The slopes of the linear fit to the data obtained are $\beta /\nu=0.249 \pm 0.004$ for $p=0.01$, $\beta /\nu=0.273 \pm 0.007$ for $p=0.02$, $\beta /\nu=0.283 \pm 0.007$ for $p=0.03$, $\beta /\nu=0.293 \pm 0.008$ for $p=0.04$ and $\beta /\nu=0.289 \pm 0.005$ for $p=0.05$. Similarly, Fig.~\ref{fig:10} shows the log-log plot of the maximum value of $C$ versus $L$ for several $p$ values. The slopes of the linear fit to the data obtained are $\alpha /\nu=0.714 \pm 0.013$ for $p=0.01$, $\alpha /\nu=0.846 \pm 0.020$ for $p=0.02$, $\alpha /\nu=0.966 \pm 0.031$ for $p=0.03$, $\alpha /\nu=1.150 \pm 0.040$ for $p=0.04$ and $\alpha /\nu=1.282 \pm 0.057$ for $p=0.05$. Similarly, Figs.~(\ref{fig:11}) and (\ref{fig:12}) show a log-log plot of the maximum values of $\chi$ versus $L$ for $p=0.01$ and $p=0.05$, respectively. The estimated values for $\gamma /\nu$ for $p=0.01$ and for $p=0.05$ are $\gamma /\nu=1.554 \pm 0.007$  and $\gamma /\nu=1.956 \pm 0.056$, respectively. For $p=0.02$, $p=0.03$ and $p=0.04$ (not shown here), the estimated values of $\gamma /\nu$ are $1.639 \pm 0.022$, $1.736 \pm 0.039$ and $1.852 \pm 0.046$, respectively.  In these figures, the statistical error bars are estimated by taking up to 100 different trial runs for each data point. In all cases studied, the estimates for $1/\nu$, $\beta /\nu$, $\alpha /\nu$ and $\gamma /\nu$ are not in the same universality class as those on a square lattice~\cite{baxter82}. The estimates of the ratios of the leading critical exponents and the average value of $\nu$ for different density $p$ are summarized in Table~\ref{tab:2}.

\section{\label{sec:c} Conclusions}

We performed Monte Carlo simulations of the three-state Potts model on NLDSW lattices to study the critical behavior presented by these systems. Both the infinite critical temperature and the leading critical exponents were estimated for different cases. Our analysis has revealed that even small values  of NLDSW disorder densities $(p\sim0.01)$ are sufficient to change the universality class when compared to those on two-dimensional periodic lattices. Furthermore, based on different quantities calculated in these paper, we claim that the three-state Potts model on NLDSW lattices exhibits a continuous phase transition for a disorder densities $p<p^{*}=0.05(4)$ falling into a new universality class, which continuously depends on density $p$, while for $p^{*}\leqslant p \leqslant 1.0$, the model presents a weak first-order phase transition.

In addition, spin-like models on NLDSW topology can be used to model real complex phenomena such as the sexually transmitted  diseases spreading and genetic recombination acting in both the DNA repair and its replication. Nodes on these lattices could represent either people or small portions of genetic material, while NLDSW links could mimic either non-parental relationships or exchanges of genetic information. We believe that both the phase transition crossover exhibited by the model studied here and its universality class might contribute to understand the general factors that rule the behavior of these systems. The former seems to be more related to internal factors such as enzymatic influences in the biological case and cultural behavior of a determined population in the social case. The later is more related to external factors such as temperature and substrate supply in the biological case and social-economic conditions in the social case.

 Finally, we expect that the results presented in this paper can be helpful to understand as non-local disorders affect the critical properties of more complex systems. Future work on NLDSW lattices will involve numerical studies on random boolean networks, spin glass and percolating systems.

\section{Acknowledgements}
We wish to thank UFERSA for computational support.


\bibliographystyle{model1a-num-names}







\end{document}